\documentclass[twocolumn,10pt,showpacs,amsmath,amssymb,prl]{revtex4}
\usepackage{graphicx}
\usepackage{dcolumn}

\begin{document}

\title{Measurement of Thermal Noise in Multilayer Coatings with Optimized Layer Thickness}
\date{\today}
\author{Akira E. Villar, Eric D. Black, Riccardo DeSalvo, Kenneth G. Libbrecht}
\affiliation{LIGO Laboratory, California Institute of Technology \\
Mail Code 264-33, Pasadena CA 91125}
\author{Christophe Michel, Nazario Morgado, Laurent Pinard}
\affiliation{Laboratoire des Mat\'{e}riaux Avanc\'{e}s, Universit\'{e} Claude Bernard Lyon 1 \\
CNRS/IN2P3, Villeurbaune, France}
\author{Innocenzo M. Pinto, Vincenzo Pierro, Vincenzo Galdi, Maria Principe, Ilaria Taurasi}
\affiliation{Waves Group, University of Sannio at Benevento \\
Benevento, Italy, INFN and LSC}

\begin{abstract}
A standard quarter-wavelength multilayer optical coating will produce the highest reflectivity for a given number of coating layers, but in general it will not yield the lowest thermal noise for a prescribed reflectivity. Coatings with the layer thicknesses optimized to minimize thermal noise could be useful in future generation interferometric gravitational wave detectors where coating thermal noise is expected to limit the sensitivity of the instrument. We present the results of direct measurements of the thermal noise of a standard quarter-wavelength coating and a low noise optimized coating. The measurements indicate a reduction in thermal noise in line with modeling predictions.
\end{abstract}	
\maketitle

\section{Introduction}
\begin{center}
\begin{figure}
\includegraphics[width=3.2in]{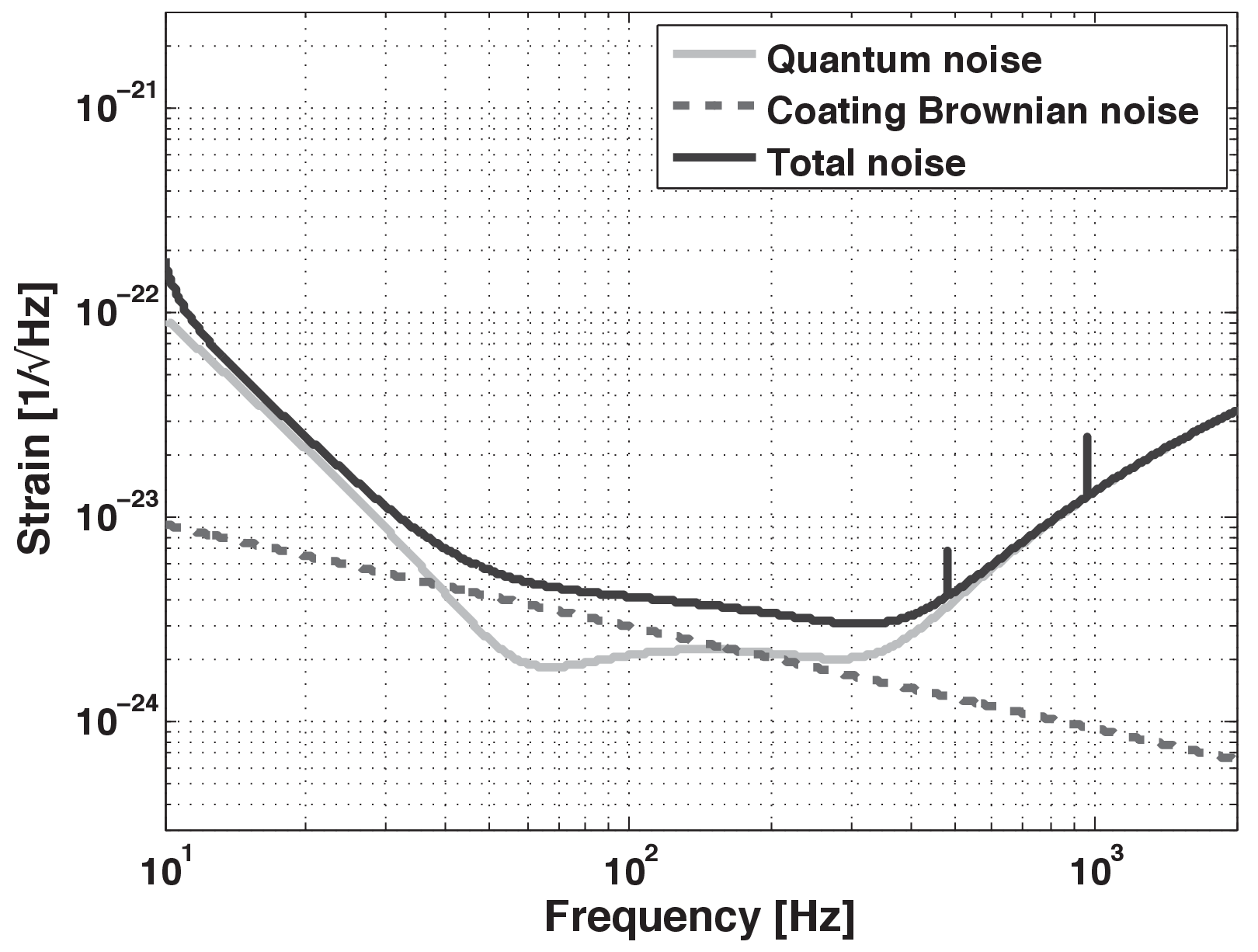}
\caption{\label{adligonoise} Projected noise floor of Advanced LIGO~\cite{gwinc}. As it stands, Brownian noise in the test mass coatings will prevent the instrument from reaching the quantum limit.}
\end{figure}
\end{center}

Several second generation gravitational wave detectors are under construction now (Advanced LIGO~\cite{Fritschel03, AdLIGODesign}, Advanced Virgo, GEO-HF~\cite{Willke06}). In general, the sensitivity of these instruments will be limited at low and high frequencies by quantum noise, however, in a critical mid-band around 100 Hz, the thermal noise of the test mass mirror coatings is a significant limit. A typical sensitivity curve for a second generation detector is shown in Fig.~\ref{adligonoise}. In light of this fact, the study of thermal noise in dielectric coatings is an active area of research in the gravitational wave community.

The optical cavities at the heart of the length-sensing mechanism of gravitational wave interferometers use mirrors made with multilayer dielectric coatings to produce the high reflectivities that are required. For good performance (low absorption) at $\lambda_0 = 1064$ nm (the operating wavelength of most gravitational wave detectors), the coatings are usually made of alternating layers of silica ($\mathrm{SiO_2}$) and tantala ($\mathrm{Ta_2 O_5}$). In an elegantly conceived experiment that studied mechanical loss in these coatings, Penn et al. \cite{Penn03} showed that the primary source of mechanical dissipation in these coatings was in the tantala layers rather than in the silica layers or at the interfaces between layers. This result suggests two possible mechanisms for reducing the mechanical dissipation and, therefore, the in-band Brownian noise~\cite{Kubo}. First, one can alter the chemistry of the tantala to reduce its mechanical loss. This has been accomplished by doping the tantala with titania ($\mathrm{TiO_2}$) to good effect \cite{Harry07}. Second, one can modify the geometry of the coating to reduce the total amount of tantala while preserving the reflectivity of the coating. Such a coating is referred to here as an ``optimized coating'' and is the subject of this paper.

A standard high-reflectivity multilayer dielectric coating consists of alternating layers of high and low index of refraction materials where the thickness of each layer is $1/4$ the local wavelength of the light. This design requires the minimum number of layers to achieve a prescribed reflectance. It does not, however, yield the lowest thermal noise for a prescribed reflectance. If there is a difference in the mechanical loss of the high and low index materials then the overall coating dissipation can be reduced by decreasing the total amount of the high loss material. We have developed a systematic procedure for designing minimal-noise coatings featuring a prescribed reflectivity. We have manufactured mirrors based on such a design at the Laboratoire des Mat\'{e}riaux Avanc\'{e}s and measured the broadband noise floor of the Thermal Noise Interferometer (TNI) at the California Institute of Technology with these mirrors installed. We then compared this to an earlier measurement of the noise floor of the TNI when mirrors with standard quarter-wavelength (QWL) coatings manufactured at Research Electro-Optics, Inc. were in place. This paper summarizes the relevant theoretical background, and presents the results of these measurements.

\section{The Thermal Noise Interferometer}

Until 2002 \cite{Numata02}, surface fluctuations of thermal origin in the mirrors of an optical cavity had never been directly observed. Up to that point the small scale of the fluctuations meant that they were always below the level of other noise sources such as laser frequency noise, shot noise, etc. The TNI is a test bed interferometer specifically designed to detect these surface fluctuations. It consists of a Lightwave Model 126 NPRO (Non-Planar Ring Oscillator) laser, a triangular mode cleaner cavity used to spatially filter and stabilize the frequency of the laser beam, and two high finesse test cavities where cavity length noise measurements are made. The key features of the TNI design that enable resolution of thermal noise are a relatively high power laser of nearly 1 W, frequency stabilization by the mode cleaner by a factor of 1000 compared to the free-running laser frequency noise in the detection band, a relatively small beam radius of 164 $\mu$m to amplify the effect of the surface fluctuations, short test cavities to reduce the laser frequency stabilization requirement, and two identical test cavities to permit common mode noise rejection. See reference~\cite{Black04} for more details about the TNI.

\begin{center}
\begin{figure}
\includegraphics[width=3.4in]{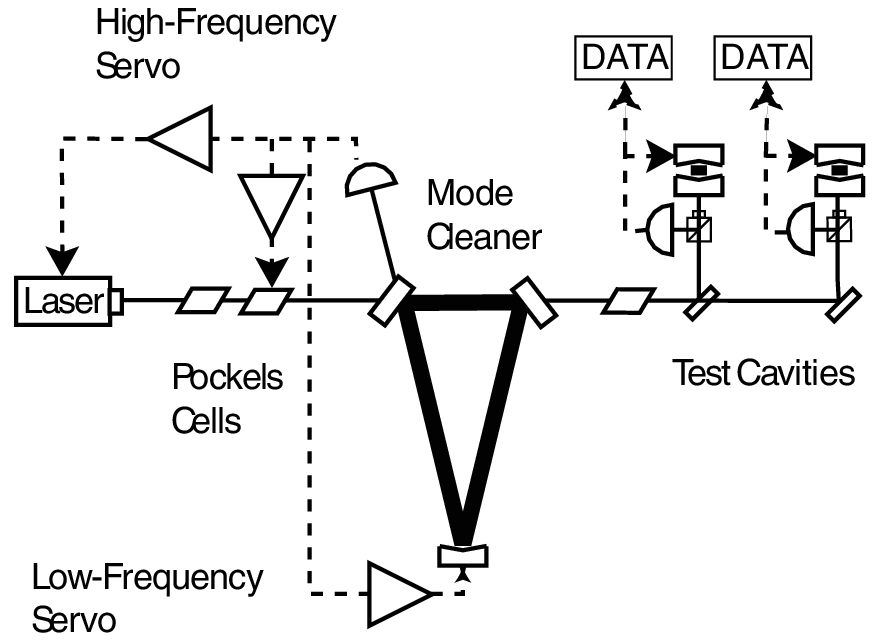}
\caption{\label{layout} Schematic of the Thermal Noise Interferometer.}
\end{figure}
\end{center}

\begin{center}
\begin{figure}
\includegraphics[width=2.8in]{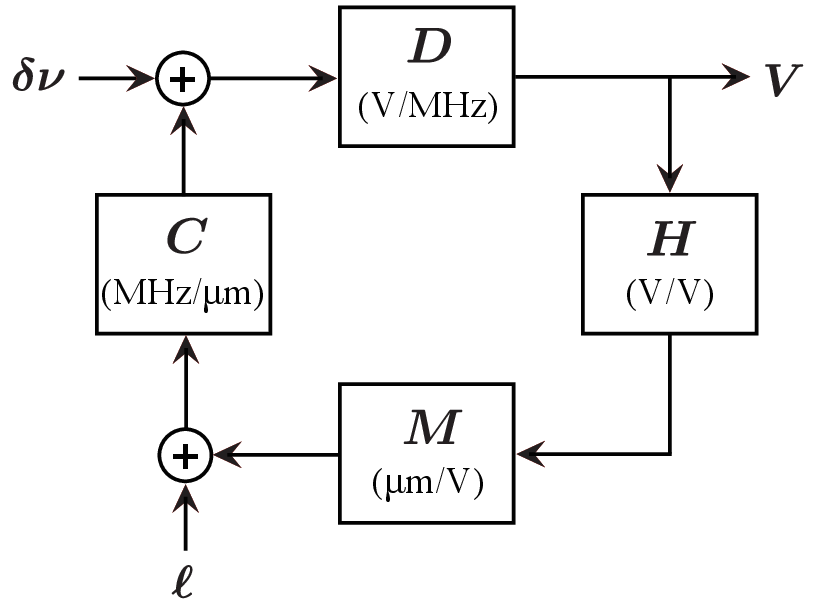}
\caption{\label{servodiagram} Block diagram of the feedback loop of a test cavity. $\ell$ and $\delta \nu$ denote fluctuations in cavity length and laser frequency, respectively. Data is taken at the point $V$.}
\end{figure}
\end{center}

The layout of the TNI is shown in Fig.~\ref{layout}. All three optical cavities are under vacuum and every cavity mirror is suspended like a pendulum from a loop of steel wire to preserve its high mechanical Q and to isolate it from external sources of noise. The Pound-Drever-Hall technique~\cite{Drever83, Black01} is used to keep the cavities locked on resonance. In the case of a test cavity, a signal is generated proportional to the deviation from the cavity resonance and this signal is fed back, after suitable amplification and filtering, to electromagnetic actuators at the end mirror of the cavity to control the cavity length. A block diagram of a test cavity servo is shown in Fig.~\ref{servodiagram}. It is a simple matter to show that a fluctuation measured at the readout point, $V$, can be converted to its equivalent fluctuation in cavity length, $\ell$, with the following formula, in the spectral (frequency) domain:

 \begin{equation}
 \ell =  T^{-1} V= \frac{1-DCMH}{DC} V.
 \end{equation}

It is imperative that the (spectral) transfer functions of the servo elements $D$, $C$, $M$, and $H$ are known accurately so that this conversion is valid. This calibration error is the largest source of uncertainty at the TNI.

\begin{center}
\begin{figure}
\includegraphics[width=2.8in]{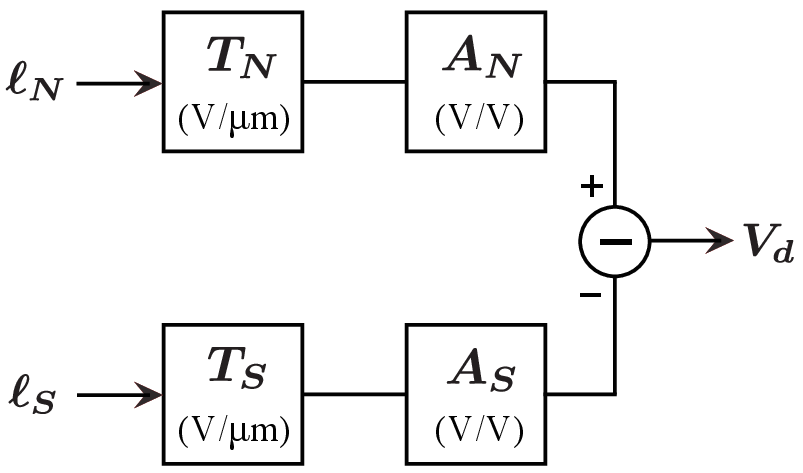}
\caption{\label{difservo} Block diagram illustrating how the difference signal readout, $V_d$, is related to the length noise in the ``North,'' $\ell_N$, and ``South,'' $\ell_S$, test cavities. $A_N$ and $A_S$ are attenuators used to balance out any difference in the responses of the servos $T_N$ and $T_S$.}
\end{figure}
\end{center}

In a single test cavity at the TNI, laser frequency noise begins to dominate the noise floor above about 6 kHz. The frequency range where thermal noise is dominant can be greatly extended if common mode rejection is implemented between the two test cavities. This is done by using a commercial pre-amplifier (Stanford Research Systems SR560) to subtract the readout signal of one test cavity from the other as shown in Fig.~\ref{difservo}. If the two test cavity servos are well matched, the difference between the readout signals, $V_d$, is related to the difference between the cavity length fluctuations in the following manner:
\[
\ell_d \equiv \ell_N - \ell_S 
\]
\vspace{-\belowdisplayskip}
\vspace{-\abovedisplayskip}
\begin{equation}
\ell_d = T_d^{-1} V_d = \frac{1}{2} \left(A_N^{-1} T_N^{-1} + A_S^{-1} T_S^{-1} \right) V_d.
\end{equation}
We refer to the two test cavities as the ``North'' and ``South'' cavities and the subscripts $N$ and $S$ reflect this. $T_N^{-1}$ and $T_S^{-1}$ are the transfer functions given by Eq. (1) for the North and South cavities, respectively. By converting the difference readout in this manner, we can compare the noise spectrum of the TNI to the theoretically predicted thermal length noise. 

At the TNI, thermal noise in the coatings can be observed for over a decade of frequencies starting at around 800 Hz. This is above the unity gain frequency of the servos (typically around 500 Hz) where the conversion formula simplifies to: 
\begin{equation}
\ell_d = \frac{1}{2 C} \left( \frac{1}{A_N D_N} + \frac{1}{A_S D_S} \right) V_d.
\end{equation}
Here we have used the fact that the servo element $C$, which is the ratio of the frequency of the laser to the length of the cavity, is the same for both test cavities. All the servo responses in Eq.~(3) are frequency independent constants. The elements $A_N$ and $A_S$ are attenuators whose gains are adjusted so that the product $A_N D_N = A_S D_S$, to maximize the common mode rejection ratio.

Once the common mode noise is removed from the TNI by subtracting the test cavity readouts from each other, the noise floor of the instrument above about 20 kHz becomes dominated by shot noise. For each test cavity, the magnitude of the shot noise is determined by the power incident on the cavity's photodetector. This is light that is reflected from the test cavity and it consists primarily of the component of the test cavity input beam that is not mode-matched to the cavity (the mode-matched component is transmitted through the cavity when it is resonant). The power depends on how well the cavities are aligned with their input beams and this can vary between measurements. The shot noise as a function of power on the photodetectors has been measured using a heat lamp as a shot noise limited source of radiation. Accordingly, determining the magnitude of shot noise at the time a thermal noise measurement is made is simply a matter of noting the power on the photodetectors at that time and using the results from the heat lamp measurement.

\section{Optimized Coating Design}

Designing an optimized coating is a matter of finding the configuration of layer thicknesses that minimizes the Brownian noise of the coating as a whole for a prescribed reflectance. Since the QWL design has the maximum reflectance for a given number of layers, the reflectance can only be maintained by adding more layers to the coating. To design the optimized coating, a means of calculating the reflectance of a coating as a function of the layer thicknesses is needed.

\begin{center}
\begin{figure}
\includegraphics[width=3.4in]{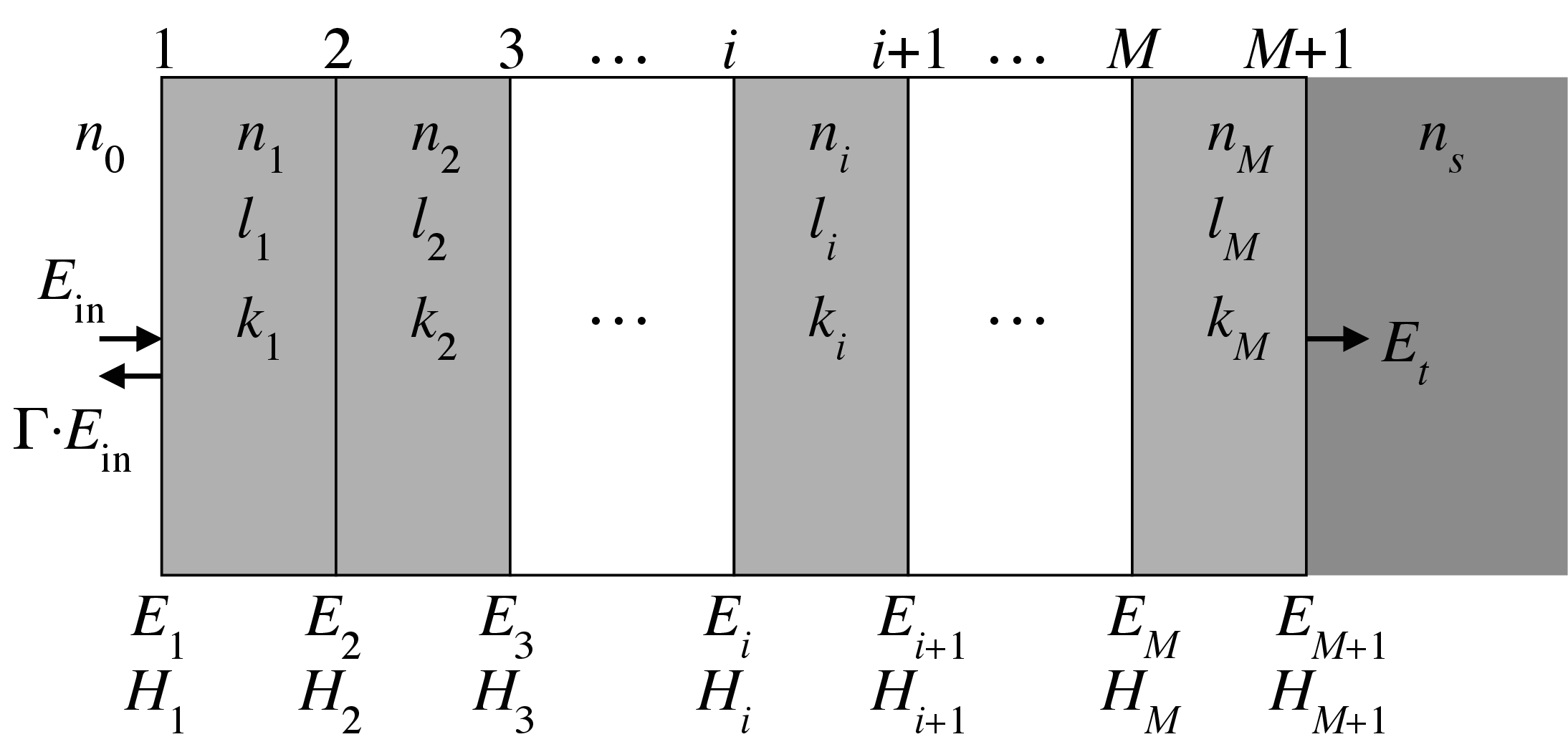}
\caption{\label{multidiagram} Diagram of a $M$ layer coating. $l_i$ is the thickness of the $i$-th layer and $n_i$ is its index of refraction. $E_i$ and $H_i$ are the electric and magnetic fields, respectively, at the $i$-th interface.}
\end{figure}
\end{center}

For time-harmonic ($e^{j \omega t}$) normal plane wave incidence, the electric and magnetic fields at the $i$-th interface of a coating are related to the fields at the $(i+1)$-th one by the propagation matrix of the $i$-th layer, $\mathbf{M}_i$ \cite{Orfanidis}.
\begin{equation}
\left[ \begin{array}{c}
E_i	\\	H_i
\end{array} \right] =
\left[ \begin{array}{cc}
\cos \delta_i	&	j n_{i}^{-1} Z_0 \sin \delta_i \\
j n_i Z_{0}^{-1} \sin \delta_i 	&	\cos \delta_i \\
\end{array} \right]
\left[ \begin{array}{c}
E_{i+1} \\ H_{i+1}
\end{array} \right]
\end{equation}
where $Z_0 = \sqrt{\mu_0 / \epsilon_0}$ is the characteristic impedance of the vacuum, $n_i$ is the refractive index of the layer, and $\delta_i = k_i l_i$ is the phase thickness of the layer. It is convenient to write the phase thickness as $\delta_i = 2 \pi z_i$, where $z_i$ is the thickness of the layer in units of the local wavelength, $z_i = l_i/\lambda_i = l_i n_i/\lambda_0$. See Fig.~\ref{multidiagram} for an illustration of a multilayer coating. The input impedance of the coating, defined as the ratio of the electric and magnetic fields at the vacuum-coating interface, $Z_{\mathrm{in}} = E_1 / H_1$, can then be obtained by chain multiplying the propagation matrices of all the layers in the coating:
\[
\left[ \begin{array}{c}
E_1  \\  H_1
\end{array} \right] = 
\mathbf{M}_1 \cdot \mathbf{M}_2 \cdot \ldots \mathbf{M}_M
\left[ \begin{array}{c}
E_{M+1}  \\  H_{M+1}
\end{array} \right]
\]
\begin{equation}
\left[ \begin{array}{c}
E_1  \\  H_1
\end{array} \right] = 
\mathbf{M}_1 \cdot \mathbf{M}_2 \cdot \ldots \mathbf{M}_M
\left[ \begin{array}{c}
1  \\  n_s / Z_0
\end{array} \right] E_t.
\end{equation}
In the last step we used the fact that the fields at the coating-substrate boundary can both be written in terms of the electric field of the transmitted beam, $E_t$. The reflection coefficient of the coating, $\Gamma$, in terms of the input impedance is:
\begin{equation}
\Gamma = \frac{Z_{\mathrm{in}} - Z_0}{Z_{\mathrm{in}} + Z_0}.
\end{equation}
The (power) reflectance of the coating is $R = | \Gamma |^2$. With this, we have a means of calculating the reflectance of a coating from the layer thicknesses. We also need a means of calculating the Brownian noise of the coating from the layer thicknesses.

The power spectral density (PSD) of the Brownian noise of a mirror is given by the following formula~\cite{Harry02}:
\begin{equation}
S_B (f) = \frac{2 k_B T}{\pi^{3/2} f} \frac{(1- \sigma_s^2)}{w Y_s} \phi_{\mathrm{eff}}
\end{equation}
where $k_B$ is Boltzmann's constant, $T$ is the temperature, $w$ is the half-width of the Gaussian laser beam, $\sigma_s$ is the Poisson ratio, and $Y_s$ the Young's modulus of the substrate. The PSD is proportional to the effective loss angle of the mirror, $\phi_{\mathrm{eff}}$, and, if the substrate has a high $Q$ factor, this is dominated by the loss angle of the coating, so that $\phi_{\mathrm{eff}} \approx \phi_c$. A precise formula for the coating loss angle is very complex~\cite{Harry02}, nonetheless, in the limit of small Poisson's ratios, it can be written as a weighted sum of the thicknesses of the materials in the coating:
\begin{equation}
\phi_c =  b_L d_L + b_H d_H ,
\end{equation}
where $d_{L,H}$ is the total thickness (in units of local wavelength) of all the low ($L$)  and high ($H$) index layers, which, in our case, are made of silica and tantala, respectively. The weighting factors are given by:
\begin{equation}
b_{L,H} = \frac{\lambda_0}{\sqrt{\pi} w} \frac{\phi_{L,H}}{n_{L,H}} \left(Y_{L,H}/Y_s+Y_s/Y_{L,H} \right) ,
\end{equation}
where, $Y$ and $\phi$ denote the Young's modulus and loss angle and the subscripts $L, H, s$ refer to silica, tantala, and substrate, respectively. Unfortunately, there is much uncertainty in the values of the loss angles for thin-film materials so the ratio $\gamma \equiv b_H/b_L$ can only be said, with confidence, to lie somewhere between 5 and 10. The optimized design was chosen among a few alternative ones yielding close-to-minimum noise, as the least sensitive to the inherent uncertainty in the value of $\gamma$ as regards the noise PSD reduction.

The problem of finding the optimal configuration of coating layer thicknesses may seem a difficult one if the thicknesses of each and every layer were allowed to be free variables. However, preliminary simulations based on genetic minimization of the coating loss angle at prescribed reflectance, using two refractive materials and no prior assumptions about coating geometry, support the conclusion that the optimal configuration is periodic and consists of a stack of identical high/low index layer pairs or ``doublets,'' with the exception of the terminal (top/bottom) layers~\cite{Agresti06}. Accordingly, in designing our optimized coatings we restrict to only these structurally simple stacked doublet configurations with tweaked terminal layers.

\begin{center}
\begin{figure}
\includegraphics[width=3.2in]{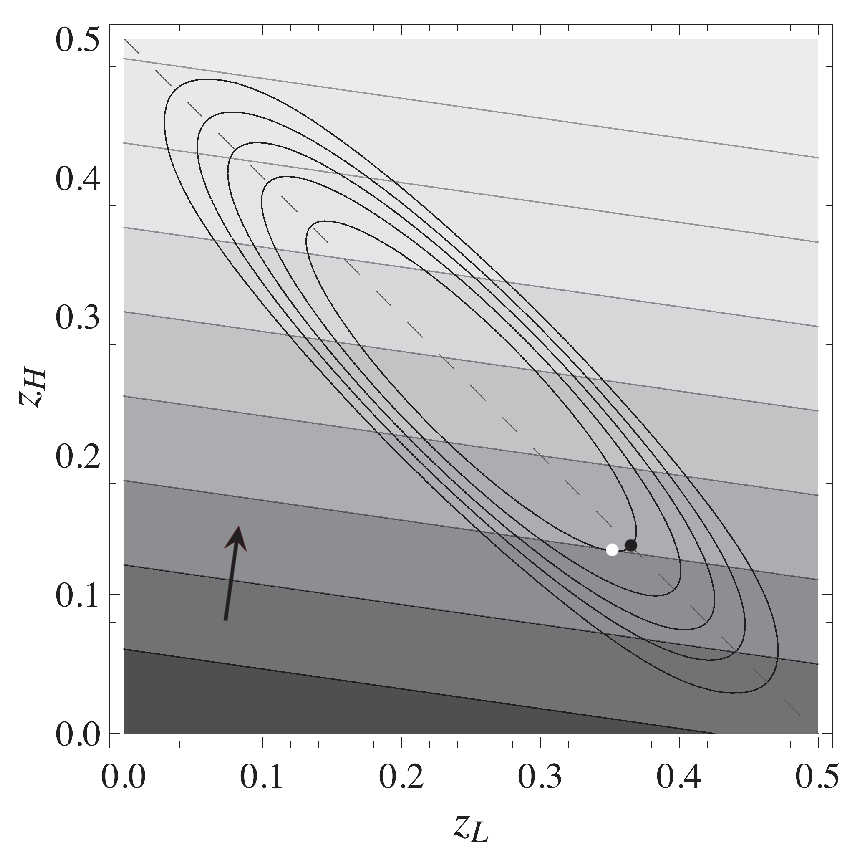}
\caption{\label{multicontours} Contours of constant reflectance (ellipse-like curves) and thermal noise (straight lines) for a stack of four identical tantala/silica doublets as functions of the silica ($z_L$) and tantala ($z_H$) layer thicknesses (in units of local wavelength). The arrow pointing from darker to lighter areas indicates the direction of increasing thermal noise. In this plot $\gamma = 7$. The dashed line consists of points where $z_L + z_H = 0.5$. The white dot marks the point of minimum thermal noise on that particular reflectance contour. The black dot marks a point on the same reflectance contour that satisfies the condition $z_L + z_H = 0.5$. This point has slightly greater thermal noise than the minimum noise point but the difference becomes negligible as the number of doublets (4 in this case) increases.}
\end{figure}
\end{center}

Further simplification is possible, though. Figure~\ref{multicontours} is a plot of contours of constant reflectance and constant thermal noise (in the case where $\gamma = 7$) as functions of the thickness (in units of local wavelength) of the silica ($z_L$) and tantala ($z_H$) layers for a coating consisting of four identical tantala/silica doublets (for a total of eight layers). In the figure, the reflectance of the coating increases to a maximum at the point $(z_L, z_H) = (0.25, 0.25)$, the QWL configuration.  The figure indicates that, for a given reflectance, there is not much difference in thermal noise between the point of minimum thermal noise (the white dot in the figure) and the point along the same reflectance contour where $z_L + z_H = 0.5$ (the black dot). Actually, as the number of doublets is increased, the reflectance contours become even more squeezed along the half-wavelength doublet dashed line and the difference in thermal noise between the above two points becomes negligible. Thus, to simplify matters further, we focus on configurations where the doublets are a half-wavelength thick. This leaves only two free parameters to adjust: the number of doublets, $N$, and the thickness of the tantala layers, $z_H$, the thickness of the silica layers being $z_L = 0.5 - z_H$. The coating design then proceeds as follows:
\begin{enumerate}
\item Starting with the standard QWL design with $N_{\mathrm{QWL}}$ doublets and reflectance $R_{\mathrm{QWL}}$, use equation (8) to calculate the loss angle of the coating.
\item Add one doublet to the coating then reduce the thickness of the tantala layers (since they are more lossy) and increase the thickness of the silica layers (to maintain the half-wavelength doublet condition) until the desired reflectance, $R_{\mathrm{QWL}}$ (calculated from Eqs. (4)--(6)), is recovered. Recalculate the loss angle of the coating.
\item Repeat step 2 and keep repeating until a design with a minimum coating loss angle, $\phi_{\mathrm{opt}}$, is obtained.
\end{enumerate}

\begin{center}
\begin{figure}
\includegraphics[width=3.3in]{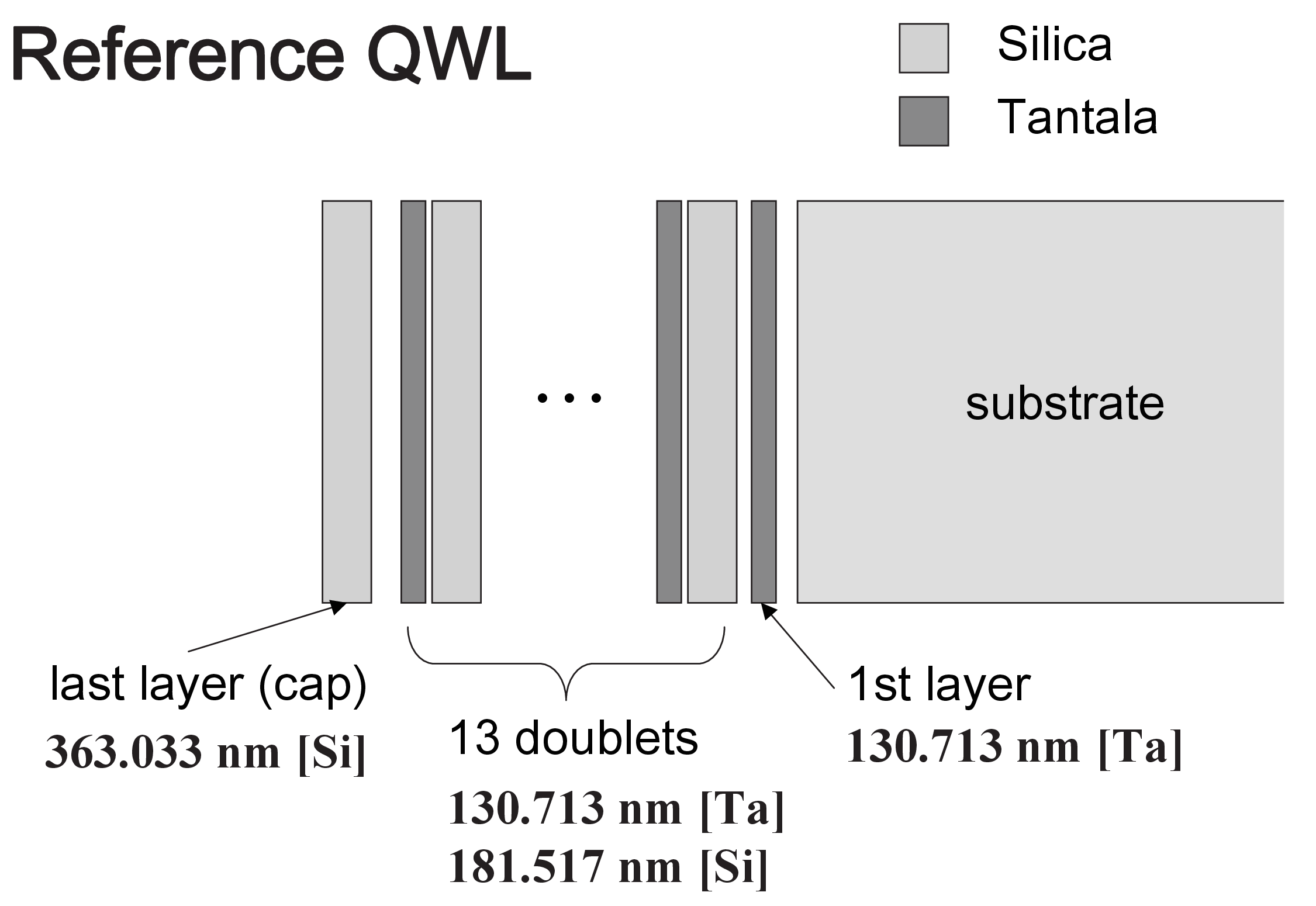}
\caption{\label{qwldesign} The structure of the standard quarter wavelength coating.}
\end{figure}
\end{center}

\begin{center}
\begin{figure}
\includegraphics[width=3.4in]{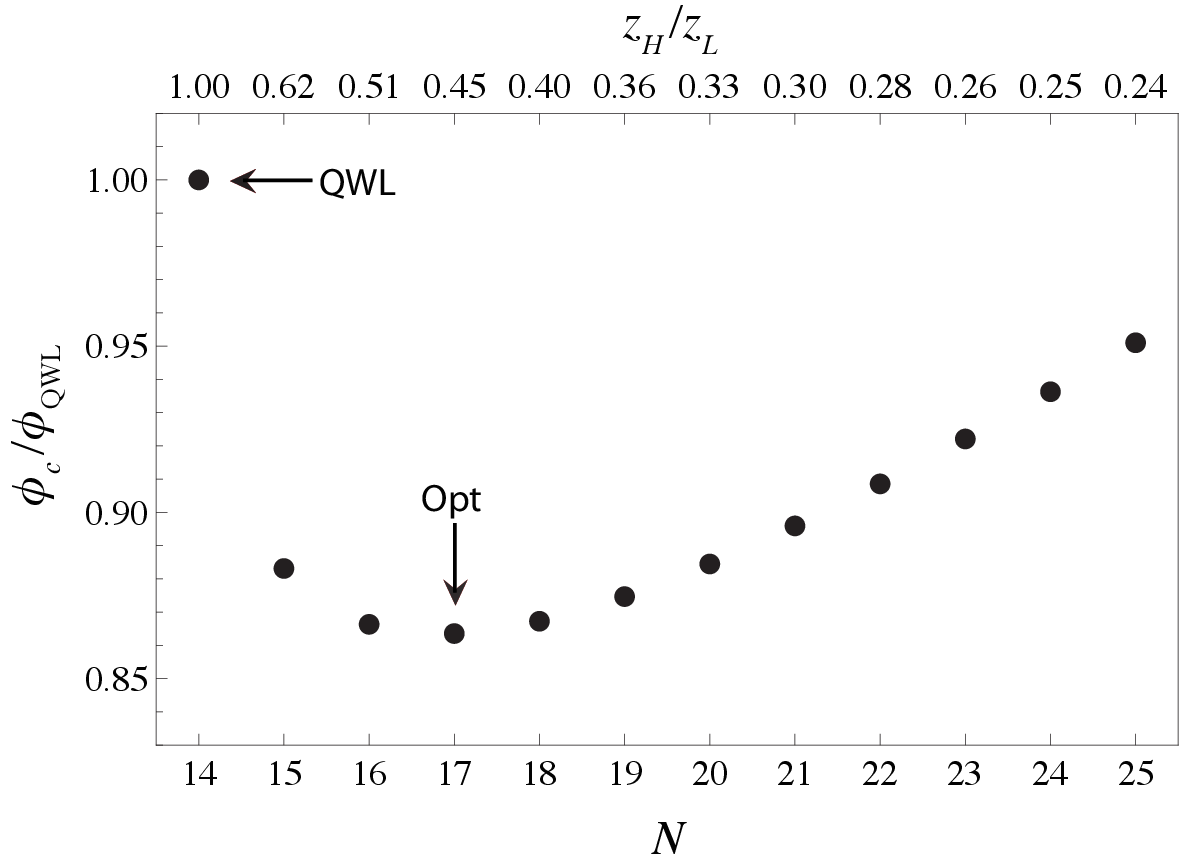}
\caption{\label{phivn} Plot of the coating loss angle (normalized by the loss angle of the QWL coating) as a function of the number of doublets, $N$. The corresponding values of the ratio $z_H/z_L$ are shown at the top of the plot. The QWL and the minimum noise (optimized) coatings are indicated in the plot.}
\end{figure}
\end{center}

\begin{center}
\begin{figure}
\includegraphics[width=3.3in]{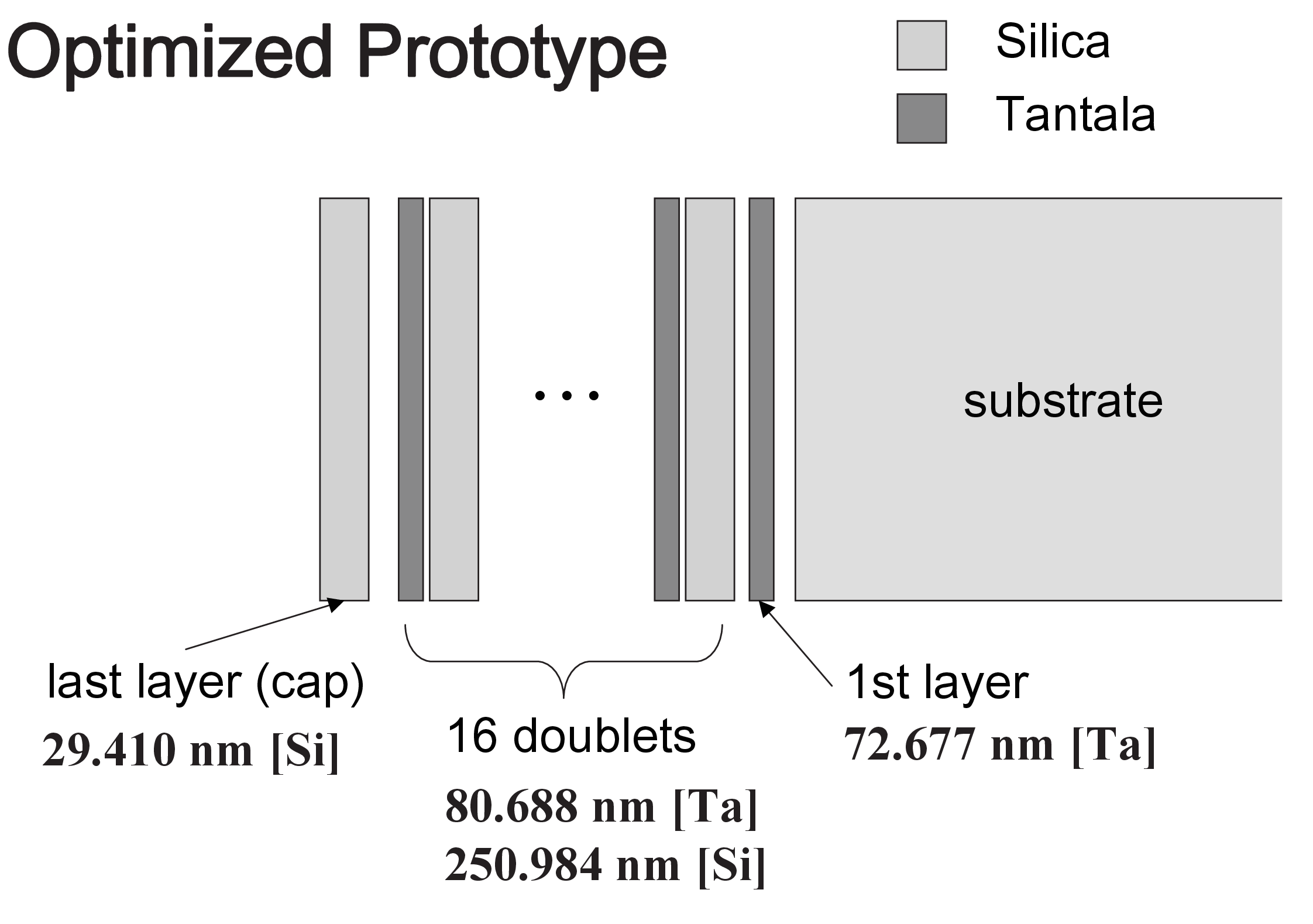}
\caption{\label{optdesign} This is the structure of the optimized coating. This coating was designed for minimal Brownian noise and a transmittance of 278 ppm.}
\end{figure}
\end{center}

The reference QWL coating measured with the TNI consists of 14 silica/tantala doublets and is shown in Fig.~\ref{qwldesign}. For protective purposes, the topmost layer of the coating is silica for its hardness. It is half-wavelength, so as to have no influence on the coating transmission properties, as seen from Eq.~(4). The reflectance of the coating is $R_{QWL} = 0.9997225$ (the transmittance is 278 ppm). Following the procedure outlined above, we designed an optimized coating with the same reflectance. The material parameters used in the design process are shown in Table~\ref{parameters}. Figure~\ref{phivn} is a plot that shows how the coating loss angle varies as the number of doublets is increased. As the figure indicates, the minimum loss angle design consists of 17 identical doublets. As suggested by the genetic simulations, this design was improved upon by tweaking the thicknesses of the end layers to further reduce the thermal noise while keeping the reflectance unchanged. The final optimized design is shown in Fig.~\ref{optdesign}. If $\gamma$ lies somewhere between 5 and 10 then the ratio of the coating loss angles, $r = \phi_{\mathrm{opt}} / \phi_{\mathrm{QWL}}$, should lie somewhere between 0.876 and 0.817. Assuming $\gamma = 7$ (the value used in the design process), then $r = 0.843$.

\begin{center}
\begin{table}
\[
\begin{tabular}{l | l}
\hline
\textbf{Coating}				& \textbf{Substrate} \\ 
\hline
$n_{\mathrm{Ta}} = 2.035 - j \ 1.8 \times10^{-7}$		& $n_s = 1.46543 - j \ 4\times10^{-8}$  		\\
$n_{\mathrm{Si}} = 1.46543 - j \ 4 \times10^{-8}$		& $Y_s = 727 \times 10^8\ \mathrm{N/m^2}$	\\
$Y_{\mathrm{Ta}} = 140 \times 10^9\ \mathrm{N/m^2}$	& $\sigma_s = 0.167$               \\
$Y_{\mathrm{Si}} = 727 \times 10^8\ \mathrm{N/m^2} $	& $\phi_s = 0$		               \\							
$\sigma_{\mathrm{Ta}} = 0.23$						& 			     	               \\
											\cline{2-2}
$\sigma_{\mathrm{Si}} = 0.167$					& $\lambda_0 = 1.064 \ \mathrm{\mu m}$               \\
$\gamma = 7$									& $T = 300$ K		\\
\hline	               
\end{tabular}
\] 
\caption{\label{parameters} The material parameters used in the design of the optimized coating.}
\end{table}
\end{center}

\section{Measurement Results}
\begin{center}
\begin{figure}
\includegraphics[width=3.4in]{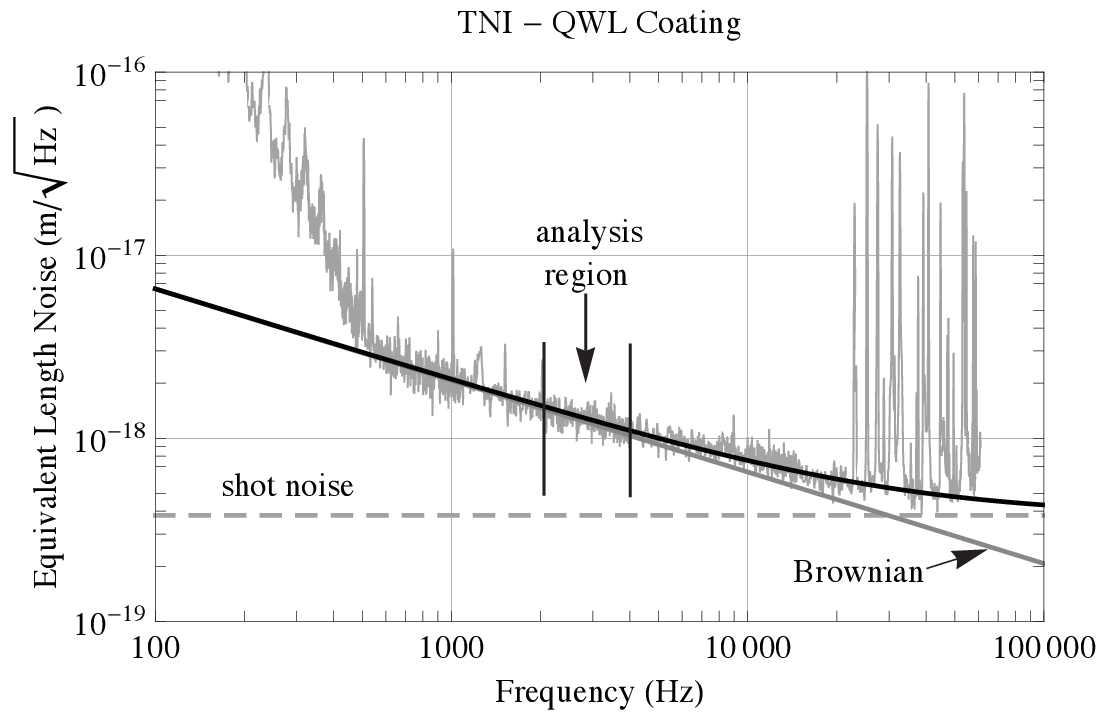}
\caption{\label{qwlspectrum}  Plot of the spectral density of cavity length noise for the QWL coating along with the shot noise (dashed line) and the Brownian noise (solid line with $f^{-1/2}$ frequency dependence). To extract the loss angle of the coating the incoherent sum of the Brownian and shot noise (solid black curve) is fit to the data between the vertical lines at 2 kHz and 4 kHz with the loss angle as the free parameter in the fit.}
\end{figure}
\end{center}

To obtain the length noise in the test cavities at the TNI, we used a spectrum analyzer (Stanford Research Systems SR780) to measure the power spectral density (PSD) at the instrument readout, $V_d$. After calibrating the instrument by measuring the responses of the servo elements in Fig.~\ref{servodiagram}, we converted this readout PSD (actually the square root of the PSD) to its equivalent length noise spectrum using Eq.~(2). The resulting length noise spectrum of the QWL coating is shown in Fig.~\ref{qwlspectrum}.  The spectrum of the optimized coating is similar. Note that from about 500 Hz to 20 kHz the measured length noise has a slope of $f^{-1/2}$ that is characteristic of Brownian noise (the square root of the PSD in Eq.~(7)). Above 20 kHz the shot noise begins to dominate the noise spectrum. The large peaks above 20 kHz are body modes of the mirror substrates. To extract the loss angles of the coatings, we focused on the region around 3 kHz since it is approximately at the center of the coating noise dominated region. 

\begin{center}
\begin{figure}
\includegraphics[width=3.4in]{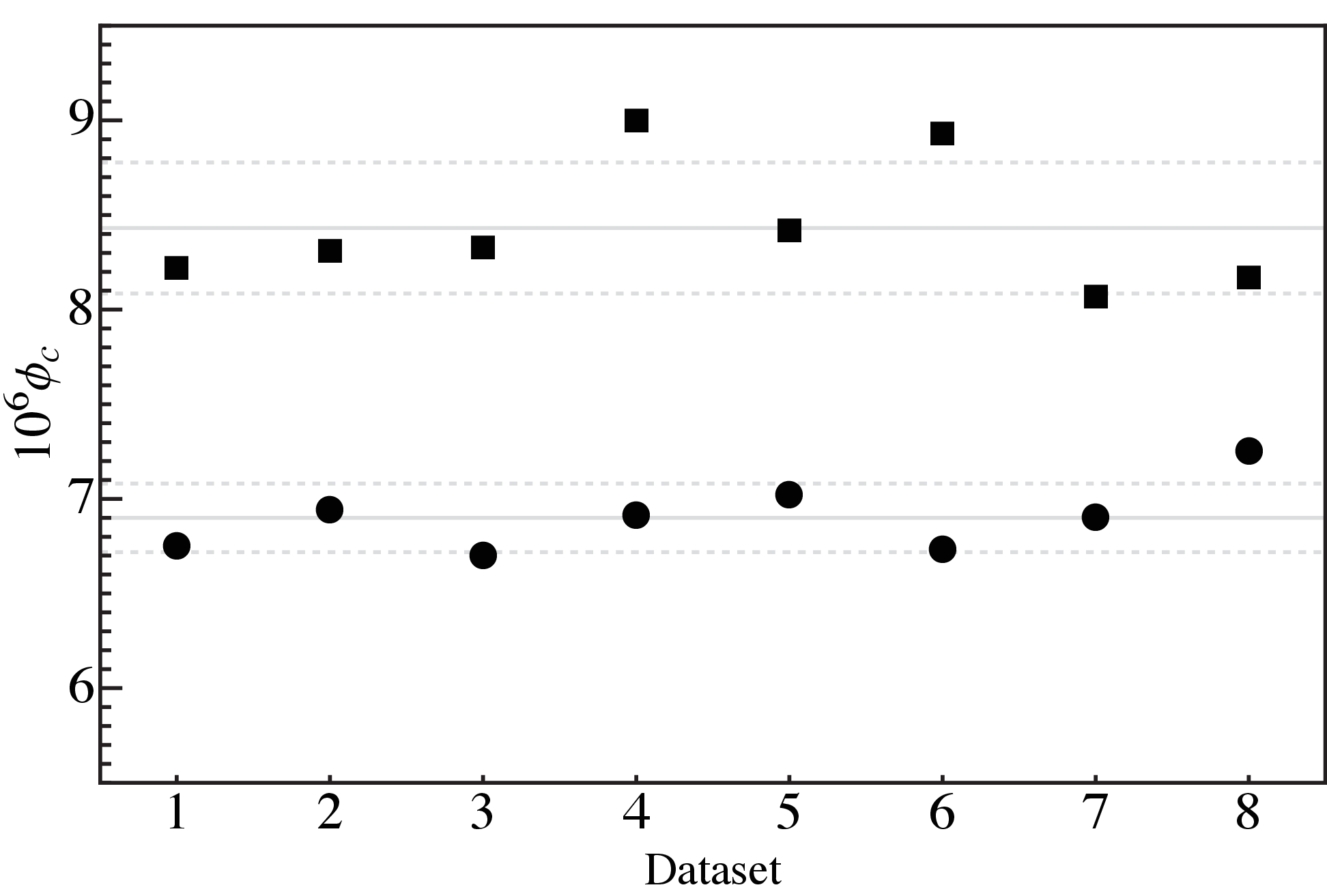}
\caption{\label{anglesplot}  Plot of the loss angles in Table~\ref{angles}. The black squares are the loss angles of the QWL coating, the black dots are the loss angles of the optimized coating. The solid gray lines are the means of each set of measurements, and the dashed gray lines are one standard deviation from the mean.}
\end{figure}
\end{center}

In the process of taking measurements of the noise floor of the instrument, we observed a certain degree of instability that is mainly due to instability of the optical gains of the servos, represented by the element $D$ in the servo diagram in Fig.~\ref{servodiagram}. To mitigate this, we took multiple measurements for each coating so that we could average the results. For each coating, we measured eight spectra from 2 to 4 kHz. Of these eight, four had a frequency bin width of 4 Hz, for a total of 500 points in each spectrum, and four had a 16 Hz bin width, for a total of 125 points in each spectrum. For each spectrum, we performed a least-squares fit of the function:
\begin{equation}
\ell(f) = \sqrt{4 \, S_B(f) + S_s}
\end{equation}
where $S_B(f)$ is the PSD of Brownian noise (Eq.~(7)), and $S_s$ is the PSD of shot noise, which is obtained from the heat lamp measurement as described earlier. The coating loss angle is the only free parameter in the fit. The factor of 4 in the equation above is due to the fact that there are a total of four mirrors in the two test cavities, so the total PSD of Brownian noise is four times that of a single mirror. The resulting loss angles are shown in Table~\ref{angles} and Fig.~\ref{anglesplot} is a graphical representation of these values. 

The mean loss angles are $\overline{\phi}_{\mathrm{QWL}} = (8.43 \pm 0.35) \times 10^{-6}$ for the QWL coating and $\overline{\phi}_{\mathrm{opt}} = (6.90 \pm 0.18) \times 10^{-6}$ for the optimized coating. To be conservative, we used the standard deviations of the two sets of eight loss angles for the errors of their respective means rather than the standard deviations divided by the square-root of eight (the number of measurements for each coating). This was done because at least some of the variation in the loss angles of each coating arises from the instability in the optical gain mentioned earlier and this variation is non-Gaussian. The ratio of the means is $r = 0.82 \pm 0.04$. Recall that the predicted ratio of optimized to QWL coating loss angles (assuming $\gamma = 7$) was 0.843, which is within the uncertainty of the measured ratio.

It is apparent from Fig.~\ref{anglesplot} that for the QWL coating, two of the measured loss angles (the fourth and sixth) are somewhat high compared to the other six. For the optimized coating, one of the values (the eighth) is high relative to the others. Each of these three values is significantly more than one standard deviation above the mean for that coating. These anomalously high measurements may be due to transients that have been observed to suddenly boost the noise floor of the instrument. They are not related to Brownian noise in the coatings. If these measurements are discarded, the means for the coatings are reduced slightly to the values indicated by asterisks in Table~\ref{angles},  $\overline{\phi}_{\mathrm{QWL}} = 8.25 \times 10^{-6}$ and $\overline{\phi}_{\mathrm{opt}} = 6.85 \times 10^{-6}$. The ratio of the means then becomes $r = 0.83$ and the remaining loss angles for each coating are more tightly grouped. While there is no compelling justification for removing these points, the new means may reflect the true coating loss angles more accurately.

\begin{center}
\begin{table}
\footnotesize
\begin{tabular}{|c|c|c|c|c|c|c|c|c||c|c|}
\hline
			& \multicolumn{4}{c|}{4 Hz bins}	&  \multicolumn{4}{c||}{16 Hz bins}  & \multicolumn{2}{c|}{}   \\
\hline
Dataset		 &  1    &   2	   &  3      &   4       &  5	  &   6        &  7   	&    8	       &  \multicolumn{2}{c|}{Mean} \\
\hline
$\phi_{QWL}$	&  8.22    & 8.31    & 8.33     & 9.00	& 8.42    & 8.93    & 8.07    & 8.17     & 8.43   & 8.25*    \\
\hline
$\phi_{opt}$	& 6.75     & 6.94    & 6.70	  & 6.91	& 7.02    & 6.73    & 6.90    & 7.25     & 6.90   & 6.85*    \\
\hline
\end{tabular}
\caption{\label{angles} Table of the coating loss angles ($\times 10^{-6}$) that give the best fit to the data along with the mean value. *Mean after discarding anomalously high values.}
\end{table}
\end{center}

\begin{center}
\begin{figure}
\includegraphics[width=3.4in]{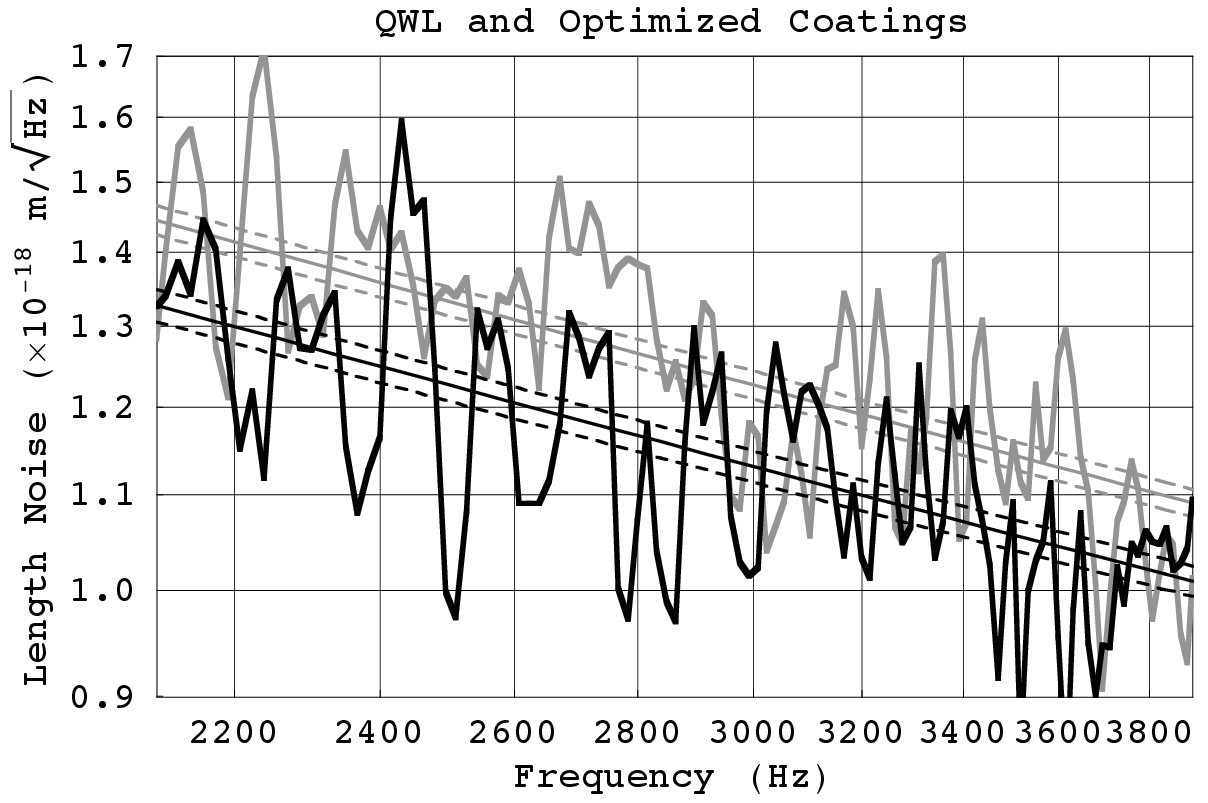}
\caption{\label{savgol0}  Plot of one of the spectra of the QWL coating (light) overlaid by one of the optimized coating (dark). The solid line through each spectrum is the best fit of Eq.~(10) to the data. For each fit, the region between the dashed lines represents the 95\% confidence interval.}
\end{figure}
\end{center}

\begin{center}
\begin{figure}
\includegraphics[width=3.4in]{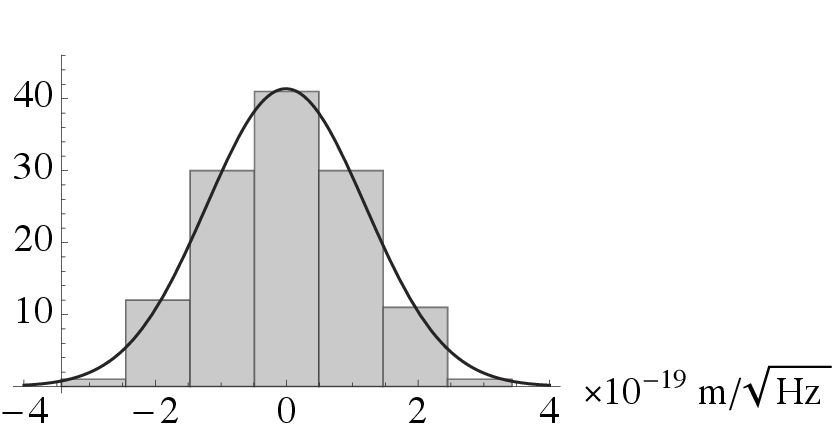}
\caption{\label{qwlhistogram} Histogram of the residuals of the fit to one of the spectra taken for the QWL coating. The distribution of the residuals is well approximated by a normal distribution with a standard deviation of $1.2 \times 10^{-19}$ m/$\sqrt{\mathrm{Hz}}$.}
\end{figure}
\end{center}

It should be noted that there is an uncertainty associated with each value in Table~\ref{angles} since each is obtained from a least-squares fit to a spectrum with some scatter in it. This is evident in Fig.~\ref{savgol0}, a plot of one of the spectra of the QWL coating overlaid by one of the optimized coating. While there is some overlap between the spectra of the two coatings there is none between the confidence intervals of the fits (indicated by the dashed lines in the plot), because the error of the fit is much smaller than the standard deviation of the residuals to the fit. Figure~\ref{qwlhistogram} is an histogram of the residuals of the fit to the QWL spectrum. The standard deviation of the histogram is $1.2 \times 10^{-19}$ m/$\sqrt{\mathrm{Hz}}$ or about 10\% of the length noise at 3 kHz and nearly equal to the separation between the spectra of the two coatings. The error of the fit, however, is smaller than the standard deviation by a factor of $1/\sqrt{N_D}$, where $N_D$ is the number of data points. Finding the least-squares fit to the data is akin to finding the center of the histogram which can be done with much more precision than its standard deviation, particularly if there are many data points. In fact, this fitting error is smaller than the errors quoted above for the mean loss angles and can be ignored. 

One more thing to note from Fig.~\ref{qwlhistogram} is that the distribution of residuals is consistent with the normal distribution, suggesting that the scatter in each spectrum is random in nature and that our method of extracting the loss angles from least-squares fits to the length noise spectra is valid.

\begin{center}
\begin{figure}
\includegraphics[width=3.4in]{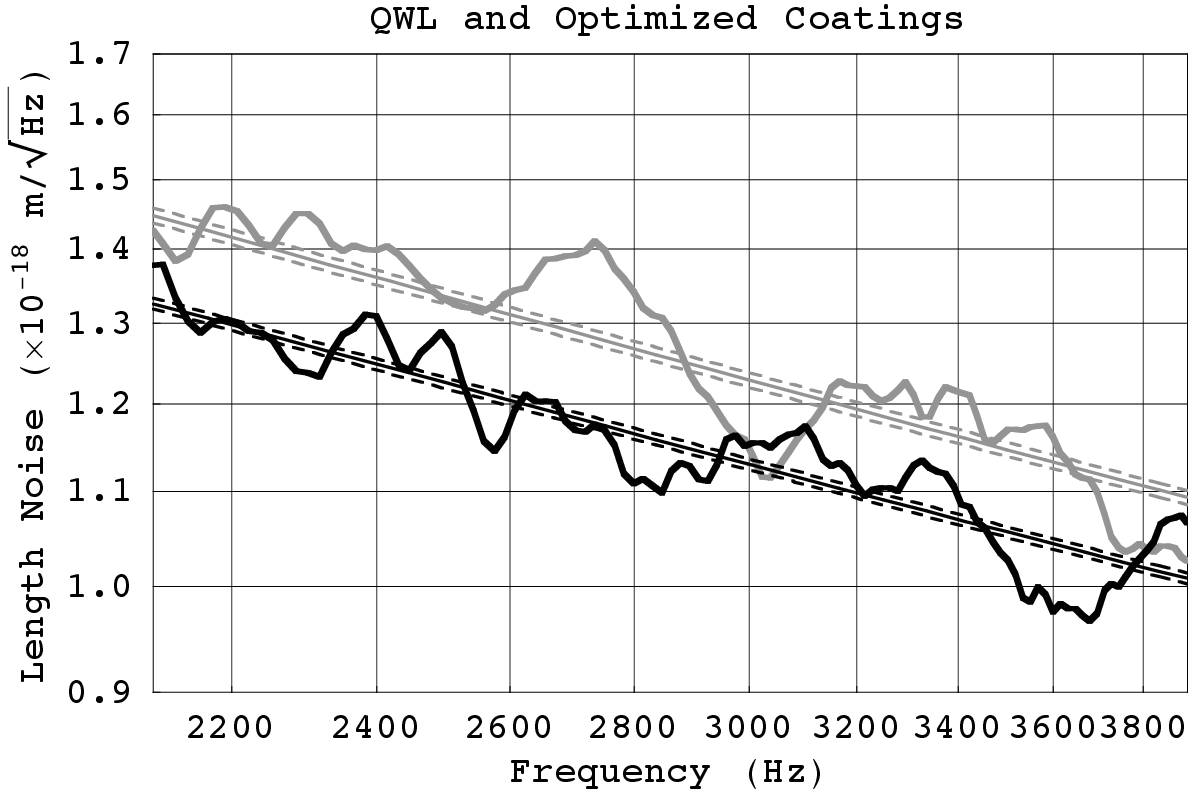}
\caption{\label{savgol5}  This is the plot in Fig.~\ref{savgol0} after the Savitzky-Golay smoothing filter was applied to each spectrum. The separation between the spectra of the two coatings can be seen more clearly.}
\end{figure}
\end{center}

The separation between the spectra of the two coatings in Fig.~\ref{savgol0} can be made visually more evident by using the Savitzky-Golay smoothing filter~\cite{SavGol} to filter out the random noise in the spectra. This technique generates a smoothed data set from the raw data where each point in the smoothed data is derived from a polynomial regression performed using neighboring points in the raw data. The smoothed data can be viewed as a generalized running average of the raw data. In the smoothed spectra shown in Fig.~\ref{savgol5}, each data point was derived from a first order polynomial regression using the corresponding point (at the same frequency) in the raw spectra of Fig.~\ref{savgol0} plus the five points to the left and the five points to the right of that point.

\section{Conclusions}
We have directly measured broadband Brownian noise of two different high reflectivity optical coating designs: a standard QWL coating, and an optimized coating specifically designed to minimize the Brownian noise. The ratio of the coating loss angles that we observed was $r = 0.82 \pm 0.04$, which agrees with the predicted ratio, to within the margin of error. The results validate the proposed coating optimization strategy, and suggest its use to improve the sensitivity of future generations of interferometric gravitational wave detectors at relatively little cost.

\section{Acknowledgments}
This research was supported by the NSF under Cooperative Agreement PHY-0757058 and the Italian National Institute for Nuclear Physics (INFN) under the Committee-V COAT grant).

\bibliographystyle{apsrev}

\end{document}